\documentclass[structabstract,longauth]{aa}
\usepackage{graphicx}
\usepackage{txfonts}
\usepackage{natbib}
\bibpunct{(}{)}{;}{a}{}{,}

\begin{document}
   \title{The Vimos VLT Deep Survey:}

   \subtitle{Stellar mass segregation and large-scale galaxy environment 
	     in the redshift range $0.2<z<1.4$\thanks{Based on data obtained with the European Southern Observatory Very Large Telescope, Paranal, Chile, program 070.A-9007(A), and on data obtained at the Canada-France-Hawaii Telescope, operated by the CNRS of France, CNRC in Canada and the University of Hawaii.}}

   \author{M. Scodeggio \inst{1}
\and D. Vergani \inst{2,1}
\and O. Cucciati \inst{3,10}
\and A. Iovino \inst{3}
\and P. Franzetti \inst{1}
\and B. Garilli \inst{1}
\and F. Lamareille \inst{4}
\and M. Bolzonella  \inst{5}
\and L. Pozzetti  \inst{5}
\and U. Abbas \inst{6,25}
\and C. Marinoni \inst{11}
\and T. Contini \inst{4}
\and D. Bottini \inst{1}
\and V. Le Brun \inst{6}
\and O. Le F\`evre \inst{6}
\and D. Maccagni \inst{1}
\and R. Scaramella \inst{7,12}
\and L. Tresse \inst{6}
\and G. Vettolani \inst{7}
\and A. Zanichelli \inst{7}
\and C. Adami \inst{6}
\and S. Arnouts \inst{13,6}
\and S. Bardelli  \inst{5}
\and A. Cappi    \inst{5}
\and S. Charlot \inst{8,14}
\and P. Ciliegi    \inst{5}
\and S. Foucaud \inst{15}
\and I. Gavignaud \inst{16}
\and L. Guzzo \inst{3}
\and O. Ilbert \inst{17}
\and H.J. McCracken \inst{8,18}
\and B. Marano     \inst{2}
\and A. Mazure \inst{6}
\and B. Meneux \inst{19,20}
\and R. Merighi   \inst{5}
\and S. Paltani \inst{21,22}
\and R. Pell\`o \inst{4}
\and A. Pollo \inst{23,24}
\and M. Radovich \inst{9}
\and G. Zamorani \inst{5}
\and E. Zucca    \inst{5}
\and M. Bondi \inst{7}
\and A. Bongiorno \inst{19}
\and J. Brinchmann \inst{26,27}
\and S. de la Torre \inst{6}
\and L. de Ravel \inst{6}
\and L. Gregorini \inst{7}
\and P. Memeo \inst{1}
\and E. Perez-Montero \inst{4}
\and Y. Mellier \inst{8,18}
\and S. Temporin \inst{28}
\and C.J. Walcher \inst{6}
          }
   \institute{INAF IASF--Milano, via Bassini 15, 20133 Milano, Italy
         \and
Universit\`a di Bologna, Dipartimento di Astronomia, Via Ranzani 1, I-40127, Bologna,
Italy
\and
INAF-Osservatorio Astronomico di Brera, Via Brera 28, I-20021, Milan, Italy
\and
Laboratoire d'Astrophysique de Toulouse-Tarbes, Universit\'e de Toulouse,
CNRS, 14 av. E. Belin, F-31400 France
\and
INAF-Osservatorio Astronomico di Bologna, Via Ranzani 1, I-40127, Bologna, Italy
\and
Laboratoire d'Astrophysique de Marseille, UMR 6110 CNRS-Universit\'e de Provence, BP8,
 F-13376 Marseille Cedex 12, France
\and
IRA-INAF, Via Gobetti 101, I-40129, Bologna, Italy
\and
Institut d'Astrophysique de Paris, UMR 7095, 98 bis Bvd Arago, F-75014, Paris, France
\and
INAF-Osservatorio Astronomico di Capodimonte, Via Moiariello 16, I-80131, Napoli, Italy
\and
Universit\'a di Milano-Bicocca, Dipartimento di Fisica, Piazza delle Scienze 3, 
I-20126, Milano, Italy
\and
Centre de Physique Th\'eorique, UMR 6207 CNRS-Universit\'e de Provence, F-13288, 
Marseille, France
\and
INAF-Osservatorio Astronomico di Roma, Via di Frascati 33, I-00040, Monte Porzio 
Catone, Italy
\and
Canada France Hawaii Telescope corporation, Mamalahoa Hwy,
Kamuela, HI-96743, USA
\and
Max Planck Institut f\"ur Astrophysik, D-85741, Garching, Germany
\and
School of Physics \& Astronomy, University of Nottingham,
University Park, Nottingham, NG72RD, UK
\and
Astrophysical Institute Potsdam, An der Sternwarte 16, D-14482, Potsdam, Germany
\and
Institute for Astronomy, 2680 Woodlawn Dr., University of Hawaii, Honolulu, Hawaii, 
96822, USA
\and
Observatoire de Paris, LERMA, 61 Avenue de l'Observatoire, F-75014, Paris, France
\and
Max Planck Institut f\"ur Extraterrestrische Physik (MPE), Giessenbachstrasse 1,
D-85748 Garching bei M\"unchen,Germany
\and
Universit\"atssternwarte M\"unchen, Scheinerstrasse 1, D-81679 M\"unchen, Germany
\and
Integral Science Data Centre, ch. d'\'Ecogia 16, CH-1290, Versoix, Switzerland
\and
Geneva Observatory, ch. des Maillettes 51, CH-1290, Sauverny, Switzerland
\and
The Andrzej Soltan Institute for Nuclear Research, ul. Hoza 69,
00-681 Warszawa, Poland
\and
Astronomical Observatory of the Jagiellonian University, ul Orla 171, PL-30-244, 
Krak{\'o}w, Poland
\and
INAF - Osservatorio Astronomico di Torino, 10025 Pino Torinese, Italy
\and
Centro de Astrof{\'{i}}sica da Universidade do Porto, Rua das Estrelas, P-4150-762, 
Porto, Portugal
\and
Leiden Observatory, Leiden University, Postbus 9513, 2300 RA, Leiden, The Netherlands
\and
Institute of Astro- and Particle Physics, Leopold-Franzens-University 
Innsbruck, Technikerstra{\ss}e 25, A-6020 Innsbruck, Austria
}
   \date{}

 
  \abstract
   {Hierarchical models of galaxy formation predict that the properties of a dark matter
halo depend on the large-scale environment surrounding the halo. As a result of this
correlation, we expect massive haloes to be present in larger number in overdense regions  
than in underdense ones. Given that a correlation exists between a galaxy stellar mass and
the hosting dark matter halo mass, the segregation in dark matter halo mass should then 
result in a segregation in the distribution of stellar mass in the galaxy population.}
   {In this work we study the distribution of galaxy stellar mass and rest-frame
optical color as a function of the large-scale galaxy distribution using the 
VLT VIMOS Deep Survey sample, in order to verify the presence of segregation in the
properties of the galaxy population.}
   {We use the VVDS redshift measurements and multi-band photometric data to derive
estimates of the stellar mass, rest-frame optical color, and of the large-scale 
galaxy density, on a scale of approximately 8 Mpc, for a sample of 5619 galaxies in 
the redshift range $0.2<z<1.4$}
   {We observe a significant mass and optical color segregation over the whole redshift
interval covered by our sample, such that the median value of the mass distribution is 
larger and the rest-frame optical color is redder in regions of high galaxy density. 
The amplitude of the mass segregation changes little with redshift, at least in the 
high stellar mass regime that we can uniformely sample over the $0.2<z<1.4$ redshift 
interval. The color segregation, instead, decreases significantly for z$>$0.7. However,
when we consider only galaxies in narrow bins of stellar mass, in order to exclude the
effects of the stellar mass segregation on the galaxy properties, we do not observe any
more any significant color segregation. }
   {}

   \keywords{Galaxies: formation - galaxies: evolution - galaxies: fundamental parameters - 
             cosmology: observations               }

   \maketitle
%

\section{Introduction}

It is well known that galaxy properties depend on the environment the
galaxies are part of. The two best and longest known examples of such an
environmental dependence are the galaxy morphology-density relation 
\citep{Dressler_80}, that describes the increasing fraction of early-type galaxies 
in the galaxy population with the increase of the local galaxy density, and the
galaxy HI content deficiency \citep{HIdef_9clust} which is observed in
cluster late-type galaxies.

More recently, with the advent of large galaxy surveys comprising samples of
tens of thousands of objects, it has become possible to study even subtler 
environmental effects on galaxy properties. Galaxy color and star formation 
history appear to be the two properties which are most strongly correlated with 
the galaxy local environment \citep{Blanton_05, Ball_08}, although an accurate
determination of what is the typical scale-length over which these effects 
are generated is still missing. Only over the last few years some agreement is 
emerging that such a scale-length must be comparable to that of clusters
of galaxies, i.e. of the order of 1 to 2 Mpc 
\citep[][hereafter Ka04; \citealt{Blanton_07}]{Kauffmann_04_env}.

\defcitealias{Kauffmann_04_env}{Ka04}

On the other hand it is also well known that physical properties of galaxies
are mostly inter-related \citep[see][for a review]{Roberts_Haynes_araa}, and that
the galaxy total stellar mass plays a significant role in determining these
properties \citep{Scodeggio_02_cube,Kauffmann_03_stellarMass}. The best known example 
of such a role is certainly the color-magnitude or color-stellar mass relations 
observed for both early and late-type galaxies, but it is by now equally clear 
that stellar mass plays an important role in shaping the star formation history of 
a galaxy \citep[see for example][]{Gavazzi_Scodeggio_96,Kauffmann_03_stellarMass,Heavens_04}.

This complex set of inter-relations among environment, galaxy properties, and
galaxy stellar mass is certainly part of the reason why the decades old
argument about which agent, between "nature" and "nurture", is the primary
driver for galaxy differential evolution, is still far from being settled.

A relatively recent addition to the debate on this subject is a scenario where
galaxy properties depend exclusively on the mass and formation history of the
dark matter halo the galaxies are formed in, but they appear to 
be correlated with the large scale environment properties purely because of
a correlation between halo properties and the large scale environment which 
surrounds the halo (such a correlation is predicted by hierarchical models, 
see for example \citealt{Mo_White_96,Sheth_Tormen_02}). This possibility, 
discussed explicitely by \citet{Abbas_Sheth_05, Abbas_Sheth_06}, is also the 
basic assumption behind the halo-model descriptions of galaxy clustering 
that have been used quite successfully in the recent past.
In particular, hierarchical models predict the ratio of massive to low mass
dark matter haloes to be larger in dense environments than it is in
under-dense ones \citep{Mo_White_96}, and simulations show that this is indeed
the case \citep{Sheth_Tormen_02,Abbas_Sheth_05}, even when the large
scale environment is defined using the standard scale length of 8 Mpc
(significantly larger than the 1--2 Mpc scale typical of galaxy clusters,
which is the scale length over which true environmental effects are expected
to be active). Since a good correlation exists between galaxy stellar mass 
and dark matter halo mass, spanning a large range of stellar
masses and galaxy types \citep[e.g.][]{Mandelbaum_06,Yang_07}, we
can directly translate the dark matter halo mass segregation prediction of
hierarchical models into a stellar mass segregation prediction.

Observationally, such a segregation has been demonstrated to exist with
reasonable certainty only in the local Universe, using SDSS data, by
\citetalias{Kauffmann_04_env}. The magnitude of this effect is rather small,
with the median stellar mass of galaxies in the densest environments probed by 
the SDSS data being only twice as large as the one of galaxies in the most 
under-dense environments (see Ka04 for the details). 
At higher redshift the effect has never been thoroughly analyzed, except for
a brief mentioning of a qualitatively similar result by \citet{Bundy_06}, 
while discussing the stellar mass function for the galaxies in 
the DEEP2 sample. Similar conclusions are being obtained by 
Bolzonella et al. (in preparation), while studying the stellar mass function 
for the galaxies in the zCOSMOS sample \citep{zCosmos}.

In this paper we use the data from the VIMOS-VLT Deep Survey \citep{VVDS} 
to extend the \citetalias{Kauffmann_04_env} result, discussing the 
observational evidence for the presence of a stellar mass segregation over the 
whole redshift interval from $z\sim0.2$ (approximately the upper limit of the SDSS 
sample) up to $z\sim1.4$ (above this redshift our sample becomes too sparse, and our 
stellar mass estimates quite uncertain), and for a possible evolution in its
strength. We also re-analyze the presence of color segregation in this redshift 
interval (a topic extensively discussed in \citealt{Cucciati_06}, 
hereafter Cu06), and discuss the correlation between mass and color segregation. 
Sect.~\ref{sec:data} of the paper briefly summarizes the data used
in this work, while Sects.~\ref{sec:mass_segregation} and \ref{sec:color_segregation}
discuss the evidence for stellar mass segregation and the connection between color
and stellar mass segregation, respectively.

\defcitealias{Cucciati_06}{Cu06}


\section{The data}
\label{sec:data}

In this work we use the observations of the VIMOS-VLT Deep Survey 
\citep[VVDS,][]{VVDS} to derive stellar masses and local galaxy densities for a
large sample of galaxies over an extended redshift interval.

The VVDS Deep is a purely magnitude limited redshift survey targeting a random
subset of a higly complete sample of galaxies in the magnitude range 
$17.5 < I_{AB} < 24.0$ (see \citealt{McCracken_F02photom} for details on the 
photometric parent sample) in the VVDS 0226-04 field (hereafter VVDS-F02). 
Spectroscopic observations with the VIMOS multi-object spectrograph were carried out 
using 1 arcsec wide slits and the LRred grism, covering the spectral range 
from 5500 to 9400 \AA, with an effective resolution of R $\sim$ 230 at 7500 \AA. 
All data have been reduced using the VIMOS Interactive Pipeline and Graphical Interface 
\citep[VIPGI,][]{vipgi_mos,vipgi_ifu}. Highly reliable redshift measurements were
obtained for 7528 objects, which corresponds to a sampling rate of
approximately 23\% of the complete parent photometric sample. In this work we
use only galaxies with redshift within the $0.2 < z < 1.4$ interval, for a
total sample of 5884 objects. Within this redshift range the spectral coverage 
of our sample is basically uniform for galaxies of all spectral types. A comparison
between the relative aboundance of different spectral types in the spectroscopic and
in the parent photometric sample shows that any bias against early-type galaxies
in the spectroscopic sample (due to the lack of emission lines, which in turn makes
the redshift estimate more difficult to obtain) is limited below the five percent
level over the whole redshift range (see the discussion in \citealt{Franzetti_07}).

Stellar mass estimates were obtained for all these objects using the GOSSIP
spectral energy distribution modeling software \citep{GOSSIP},
taking advantage of the multi-band photometric observations available in the
VVDS-F02 field, including BVRI data from the CFHT \citep{McCracken_F02photom},
U-band data from the ESO-MPI 2.2m telescope \citep{Radovich_F02photom}, ubvrz data
from the CFHT Legacy Survey (McCracken et al. 2007), J and Ks-band data from
SOFI at the NTT \citep{Iovino_F02photom,Temporin_F02infrared} and from the UKIDSS
survey \citep{Lawrence_07}, 3.6 micron data from the Spitzer-IRAC SWIRE
survey \citep{SWIRE}. The photometric and spectroscopic data were
fitted with a grid of stellar population models, generated using the PEGASE2
population synthesis code \citep{PEGASE}, assuming a set
of "delayed" star formation histories (see \citealt{Virgo_spec} for details),
and a \citet{Salpeter_imf} initial mass function.
Further details on the derivation of the stellar masses are presented in
\citet[][see also \citealt{Pozzetti_VVDS_MF}]{Vergani_08}. 

\begin{table}
\caption{Sample General Properties}             
\label{tab:summary}      
\centering                          
\begin{tabular}{c c c c}        
\hline\hline                 
redshift bin & median z & Log($M_{limit}/M_\odot$) & N($>M_{limit}$) \\ 
\hline                        
  0.2--0.5 & 0.40 & 9.0 & 722 \\      
  0.5--0.7 & 0.61 & 9.5 & 878 \\
  0.7--0.9 & 0.82 & 9.8 & 844 \\
  0.9--1.4 & 1.12 & 10.3 & 692 \\
\hline                                   
\end{tabular}
\end{table}

Local galaxy densities were obtained computing the three-dimensional number
density contrast for galaxies above a certain luminosity threshold in the
spectroscopic sample, within a fixed comoving volume. The point-like galaxy
distribution was smoothed with a Gaussian filter with a sigma of 5 Mpc,
roughly equivalent in volume to a top-hat spherical filter with a sphere
radius of 8 Mpc. Further details on the estimation of the local galaxy density
are given in \citetalias{Cucciati_06}. The choice of this smoothing length
has been dictated by the sample properties, mostly by the fact that the mean 
inter-galaxy separation typical of the VVDS Deep sample is of approximately 
4.5 Mpc at the peak of the sample redshift distribution, and even larger at 
the low and high redshift ends of the distribution. The relatively large value
for the smoothing length has also the advantage of mitigating the effects of
redshift-space distorsions created by galaxy peculiar motions in overdense 
regions on the density estimates \citepalias[see Figure 2 in][]{Cucciati_06}.

To avoid using very uncertain density contrast estimates we have removed from 
the sample objects that are located at the edges of the volume sampled by the 
VVDS Deep, for which only one third or less of the comoving volume used to 
sample the galaxy density contrast is effectively inside the survey volume. 
Therefore our final sample is composed of 5619 galaxies.

\section{Stellar mass segregation}
\label{sec:mass_segregation}

If a correlation exists between stellar mass and dark matter halo mass on one
side, and the large scale environment properties on the other side, we can
expect the strength of the observed stellar mass segregation to depend on the
range of large-scale galaxy densities being considered. Indeed the results
presented by \citetalias{Kauffmann_04_env} for the SDSS sample show exactly a 
progressive increase in the median stellar mass for samples of galaxies that range 
from isolated objects to objects in high density environments (see their Fig. 3). The
global change in median mass is small, only a factor of 2 when going from the
lowermost to the highest density sample, but nonetheless the large statistical
sample provided by the SDSS dataset makes this a very robust result.

%
   \begin{figure}
   \centering
   \includegraphics[width=9cm]{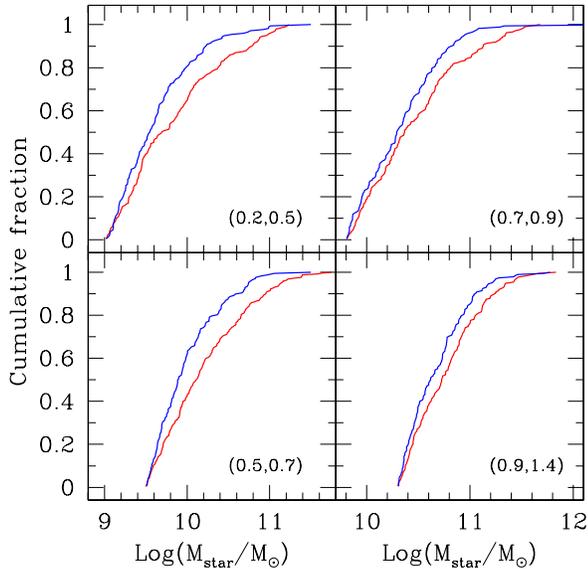}
      \caption{The cumulative mass distribution for the galaxies in the
	lower (blue line) and upper (red line) quartile of the overall 
	density distribution in four redshift bins, identified at the
	bottom of each plot; only galaxies above the stellar mass completeness 
	limit appropriate for their redshift interval are considered. }
         \label{fig:mass_integral}
   \end{figure}
%

To extend this result to higher redshift using the VVDS Deep sample,
we have to deal with the fact that the magnitude limited nature of the sample 
produces a stellar mass completeness limit which varies significantly
with redshift. As the stellar mass is correlated with galaxy color, and 
galaxy color is correlated with environment, the only meaningful way we have 
to search for a possible mass segregation effect in our data is to make sure 
we are using a mass complete sample at all redshifts. Some discussion about 
the mass completeness of the VVDS Deep sample has already been presented in 
\citet{Pozzetti_VVDS_MF}, and was further extended in \citet{Vergani_08} and 
in \citet[][see their Figure 3)]{Meneux_08} to better keep into account the 
differential mass limit as a function of galaxy color in our sample. 
Using these latter results, we have divided our samples into 4 redshift bins, 
each one with its own stellar mass completeness limit, as listed 
in Table~\ref{tab:summary}.
The table lists the limits adopted for the different redshift bins, the median
redshift for the objects in the stellar mass complete sample, the logarithm of 
the stellar mass completeness limit, in solar mass units, and the number of
objects in such a sample. Following the results presented by \citet{Meneux_08}
on the stellar mass completeness, and the discussion presented by \citet{Franzetti_07} 
on the uniformity of the VVDS spectral coverage as a function of spectral type, 
we can be confident that any color or mass-dependent incompleteness in these 4 
samples is limited below the 5 percent level, and it cannot therefore have any
significant impact on the results presented below.

Because of the relatively small number of objects within any redshift bin we 
are considering, we can only partition the sample in a limited number of 
large-scale environments. In this work we consider a partition of the various 
environments according to the estimated galaxy number density contrast. 
For the total VVDS Deep galaxy sample described in
Section~\ref{sec:data} we have obtained the distribution of the density
contrast values, and for this work, unless stated differently, we consider as 
galaxies in low density environment those objects for which the density contrast 
value is in the lower quartile of the distribution. Conversely, we consider 
galaxies in high density environment those objects for which the density contrast 
value is in the upper quartile of the distribution. As already discussed in
\citetalias{Cucciati_06}, the separation between these two extremes of the 
density distribution is very robust when using the smoothing length of
approximately 8 Mpc which is used in this work.

We find that a significant stellar mass segregation is present throughout the
VVDS Deep sample used here, up to a redshift of 1.4 (the median z value for our
highest redshift bin being 1.12).  Fig.~\ref{fig:mass_integral} shows
such a segregation, plotting the integral distribution of stellar mass values 
for galaxies in the four redshift intervals listed in Table~\ref{tab:summary}, 
limited to the mass complete samples of objects with $Log~M_{star} > Log~M_{limit}$; 
this is the analogous of Figure 3 in \citetalias{Kauffmann_04_env}. 
It is quite clear from this plot how the stellar mass distribution in the high density 
environments is skewed towards higher masses with respect to the distribution in the 
low density environments. The statistical significance of the observed segregation is 
at the three-sigma level or above for the three lower redshift bins, while it is only 
at the two-sigma level for the highest redshift bin, partly because of the lower number 
of objects we have in that bin.

%
   \begin{figure}
   \centering
   \includegraphics[width=8cm]{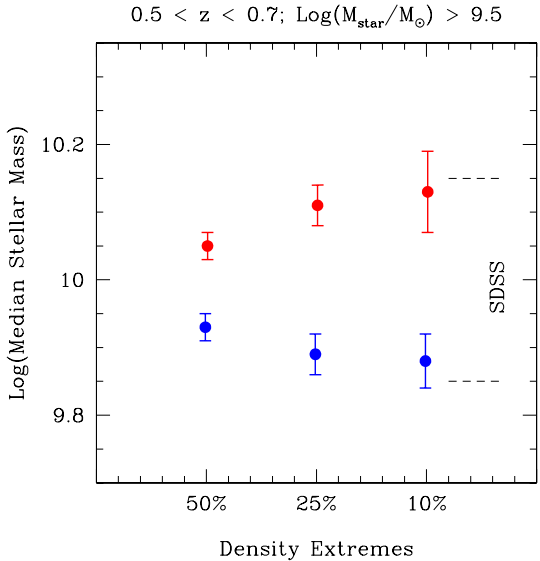}
      \caption{Median values of the mass distribution for galaxies in the 
	lower (blue points) and upper (red points) extreme of the overall 
	density distribution, for three different definitions of extreme:
	upper and lower half of the distribution, or 50\% extreme; upper and
	lower quartile of the distribution, or 25\% extreme, upper and lower
	10\% of the distribution. Only galaxies in the redshift bin $0.5<z<0.7$
	and above the mass completeness limit Log($M_{limit}/M_\odot$)=9.5 are
	considered in this plot. The error bars represent the bootstrap-based 
	uncertainty estimate in the median value determination. As a reference, the
	amplitude of the median stellar mass offset measured in the SDSS sample
	is given by the two dashed lines to the right of the plot (see text for
	details).}
         \label{fig:mass_contrast}
   \end{figure}
%

As expected, given the \citetalias{Kauffmann_04_env} results, we observe that 
the strength of this segregation depends mildly on how much the low and high 
density environments are differentiated. In Fig.~\ref{fig:mass_contrast} 
we show, limited to the redshift bin $0.5<z<0.7$, the median values for the stellar 
mass distribution for galaxies in low and high density environments as a function 
of how extreme we take these two environments to be. Therefore in this case we do not 
consider just objects in the lowermost or highest quartile (i.e. 25\% extremes) of the 
galaxy density contrast distribution, but also those in the lower and upper half 
(i.e. 50\% extreme) of the distribution, and those in the lower and upper 10\% of the 
distribution (10\% extremes). Together with the median values we also plot the 
associated uncertainties, derived using a bootstrap procedure.
Although the separation of the median stellar mass values
increases as we move towards the extremes of the density distribution
(i.e. towards the extremes of the environment), the statistical significance
of these differences remains quite constant, according to a two-populations
Kolmogorov-Smirnov test applied to the full stellar mass distributions,
because of the decreasing number of objects in the more extreme samples.
Therefore the significance with which we measure mass segregation between low
and high density environments is always approximately 3 sigmas (i.e. the
probability that the two stellar mass distributions are actually drawn from
the same parent population is always in the 0.1--0.5 percent range). 
Similar results are obtained for the other redshift bins we are considering in
this work. Purely as a reference, the offset in median stellar mass measured by 
\citetalias{Kauffmann_04_env} in the two most extreme environments they sampled is 
indicated in the figure by the two dashed lines on the right of the plot 
(the mass values where the two lines are drawn are totally arbitrary), although
we must remark that the mass completeness limit for that sample, and the definition
of environmental densities are different from those used in this work.

%
   \begin{figure}
   \centering
   \includegraphics[width=8cm]{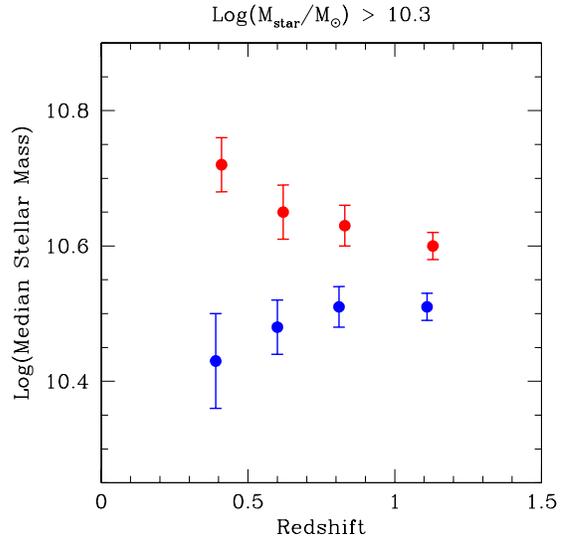}
      \caption{Median values of the mass distribution for galaxies in the 
	lower (blue points) and upper (red points) quartiles of the overall 
	density distribution, in the 4 redshift bins listed in Table~\ref{tab:summary}.
	Only galaxies above the common mass completeness limit 
	Log($M_{limit}/M_\odot$)=10.2 are considered. The error bars represent
	the bootstrap-based uncertainty estimate in the median value determination.}
         \label{fig:mass_z}
   \end{figure}
%

Finally, to examine in detail if and how this mass segregation evolves with
cosmic time, we have defined a subsample of 1872 galaxies, limited to objects
with Log$M_{star}/M_\odot > 10.3$, which is complete in stellar mass over the full
redshift range sampled by our data. Figure~\ref{fig:mass_z} shows the median
stellar mass value for galaxies in low and high density environments, over
four redshift bins (those listed in Table~\ref{tab:summary}). As in the previous 
figure, we also plot the uncertainty in the median estimates. Although we
observe a small increase in the strength of the mass segregation (as measured
from the median value of the mass distribution), from redshift of approximately 
1.1 down to redshift of approximately 0.4, this change is not statistically 
significant: if we compare any two stellar mass distributions for the same environment,
but in different redshift bins, we find a probability of more than 50 percent for 
the two to be drawn from the same parent population (according to a two-populations 
Kolmogorov-Smirnov test). This is true for both the low and the high density 
environment samples. A significantly larger mass-complete sample of galaxies
is needed before we can reliably detect any significant evolution in the strength 
of the stellar mass segregation from the local Universe to z=1.

%
   \begin{figure}
   \centering
   \includegraphics[width=9cm]{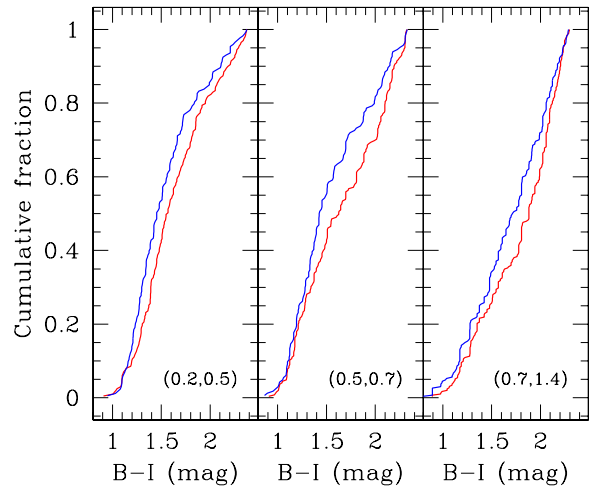}
      \caption{The cumulative rest-frame B-I color distribution for the 
	galaxies in the lower (blue line) and upper (red line) quartile of 
	the overall density distribution in the redshift bins $0.2<z<0.5$, $0.5<z<0.7$
	and $0.7<z<1.4$ respectively.}
         \label{fig:color_dist}
   \end{figure}
%

\section{Rest-frame color segregation}
\label{sec:color_segregation}

Physical properties of galaxies (like their color, dust and gas content,
and star formation activity) and their morphologies are significantly 
inter-related, and it is well documented that the galaxy total stellar mass 
plays some important role in determining these properties 
\citep[see for example][]{Scodeggio_02_cube,Kauffmann_03_stellarMass}. 
It is therefore quite natural to expect that some segregation in galaxy 
properties should be observed, purely as a consequence of the stellar mass 
segregation we have discussed in the previous section.

Rest-frame galaxy colors are well known to show such a segregation in the
local Universe \citep[see for example][]{Blanton_05,Baldry_06}, 
mirroring the equally well-know morphology-density relation \citep{Dressler_80}. 
Moving to higher redshift, both the VVDS and the DEEP2 survey have shown that 
a significant correlation between galaxy color and the large-scale environment 
exists at least up to z$\simeq$1 (see \citetalias{Cucciati_06} and 
\citealt{Cooper_06}, respectively), notwithstanding the general trend towards 
bluer colors which is observed when moving from the local Universe to z$\simeq$1. 
In particular \citetalias{Cucciati_06}, by examining the fraction of red and blue 
galaxies in different large-scale environments sampled by the VVDS dataset, have shown 
that, up to $z<0.9$, one observes the locally well known correlation between the fraction 
of red galaxies and the large-scale density of galaxies. At higher redshift 
($0.9<z<1.5$) the correlation basically disappears, as more and more massive galaxies 
become actively star-forming, and it is even possible that the correlation reverses
completely, with a predominance of blue galaxies in high-density environments
at z$\simeq$1.5.

%
   \begin{figure*}
   \centering
   \includegraphics[width=18cm]{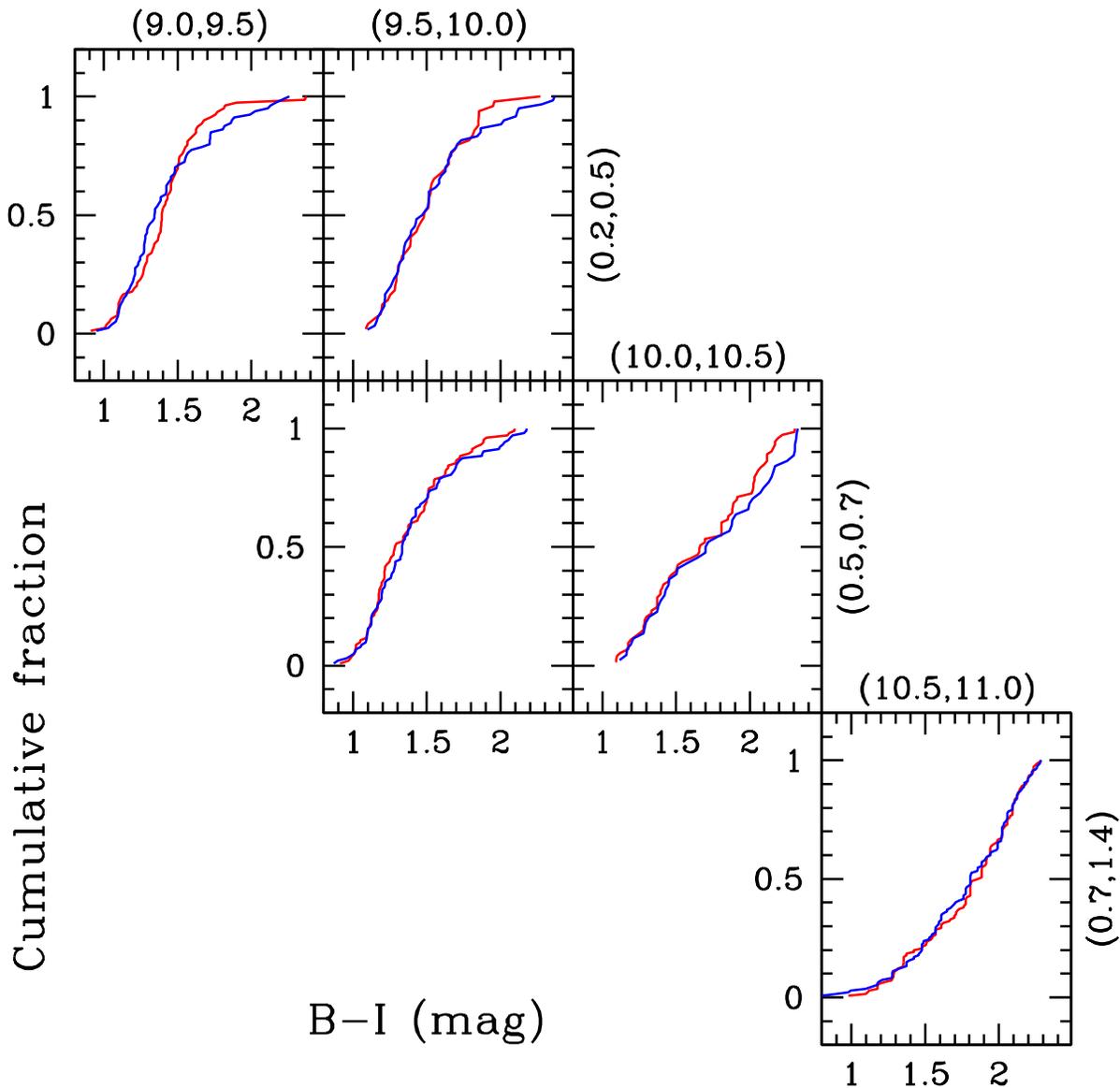}
      \caption{The cumulative rest-frame B-I color distribution for the 
	galaxies in the lower (blue line) and upper (red line) quartile of 
	the overall density distribution in the same redshift bins used in 
	the previous figure (as indicated to the right of the panels), 
	but considering only galaxies in small stellar mass bins (indicated 
	at the top of the various panels).}
         \label{fig:color_massBin}
   \end{figure*}
%

In Fig.~\ref{fig:color_dist} we confirm and extend the \citetalias{Cucciati_06}
result, by showing the presence of a significant rest-frame color segregation 
in the VVDS dataset, not only for volume-limited datasets, but also for mass complete
ones. Here we plot the integral distributions of rest-frame B-I color 
values for galaxies in the first two redshift intervals listed in Table~\ref{tab:summary}
separtately, and for galaxies in the last two redshift intervals grouped together. 
Within each plot, the whole stellar mass complete sample of objects is considered. 
Here again, because of the completeness in stellar mass and in spectral type coverage
already discussed in Section~\ref{sec:mass_segregation}, we can confidently state that
the color distributions for objects in low and high density environments are 
statistically different at the 3-sigma level (having a probability of being drawn 
from the same parent population of 0.2--0.4 percent, as estimated from a 
two-populations Kolmogorov-Smirnov test) for z$<$0.7, while in the redshift 
interval 0.7$<$z$<$1.4 the significance drops to approximately the 2-sigma level. 
This result is in complete agreement with the earlier findings of 
\citetalias{Cucciati_06} about the disappearance of the correlation between the 
fraction of red galaxies and the large-scale density of galaxies for z$>$0.9.

In Fig.~\ref{fig:color_massBin}, on the contrary, we show that the rest-frame color
segregation is significantly weaker when we consider only galaxies within relatively
small stellar mass bins, to the point that we do not observe any statistically
significant difference in the color distribution for galaxies in the low and high 
density environment. This lack of any residual color segregation is in contrast with 
some findings from the SDSS sample discussed in \citetalias{Kauffmann_04_env} and 
in \citet{Baldry_06}, that report a significant environmental dependence of galaxy 
properties, even when considering only objects within relatively small stellar mass 
bins. One possible explanation for this discrepancy is the very different scale-length 
over which the local galaxy density is estimated in this work (8 Mpc) and in the 
SDSS ones (1 Mpc). Another important difference coming into play is the smaller size 
of the VVDS Deep sample used here, with respect to the SDSS one. 
It is therefore entirely possible that some segregation in galaxy rest-frame color could 
still be present in our sample, even when considering narrow mass bins, but the relatively 
minor strength of this segregation, coupled with the small sample size, could prevent us 
from measuring this effect with any statistical significance. This topic will be analyzed 
in further detail using the larger zCOSMOS sample in a forthcoming paper 
(Cucciati et al., in preparation). 

This discrepancy aside, the observed significant weakening of the galaxy color 
segregation which is observed when we consider galaxies in relatively narrow stellar 
mass bins strongly suggests that the segregation we observe on 8 Mpc scales is mostly 
(if not entirely) driven by the underlying stellar mass segregation, coupled with 
the well known color-stellar mass correlation. A rather similar conclusion could 
be derived from the observed absence of any environmental effects on the locus of 
the color-stellar mass relation, both for objects in the red sequence and in the 
blue cloud, which was discussed by \citet{Cassata_07}, using 
the data from the COSMOS survey \citep{COSMOS}.

\section{Conclusions}

A significant stellar mass segregation as a function of the large-scale galaxy
environment, the latter defined from the measured galaxy number density contrast 
within scales of approximately 8 Mpc, is observed in the VVDS sample over the whole 
redshift range from z=0.2 up to z$\sim$1.4: the stellar mass distribution of high 
density regions shows the presence of a larger number of high stellar mass galaxies 
than the distribution observed in low density regions. The scales over which this
segregation is observed are much bigger than the typical group or cluster scale
(approximately 1 Mpc) where environmental effects are expected to play a significant 
role in shaping galaxy evolution. It is however impossible for us, with the present
sample, to evaluate the possibility that the mass segregation signature we observe 
on the 8 Mpc scale could just be a diluted signal produced by a much stronger 
segregation at the 1 Mpc scale, because the typical mean interparticle separation 
of the VVDS Deep sample does not allow us to reliably derive galaxy densities on the
1 Mpc scale. It is remarkable however that the observed segregation
is in rather good agreement with the expectations from the hierarchical models, 
which predict a very similar segregation of dark matter haloes (the strength
in the mass segregation of the dark matter haloes is expected to be 2--3 times stronger 
than the observed stellar mass one, but part of this discrepancy could be explained 
by the presence of a large number of very low mass halos in the underdense
regions that do not host any galaxy, see \citealt{Abbas_Sheth_05}). 
The strength of the observed stellar mass segregation decreases marginally with
increasing redshift, from z=0.4 to z=1.2, but within the relatively small statistics
offered by the mass complete VVDS galaxy sample such a weakening of the mass
segregation cannot be considered as a statistically robust result.

A significant rest-frame galaxy color segregation is observed as well, mirroring the 
stellar mass one.  However we have demonstrated that a large fraction of this color 
segregation is simply a reflection of the stellar mass segregation, via the well
known correlation between stellar mass and galaxy color. In fact, when we compare
rest-frame colors for galaxies with stellar mass within a narrow range, we do not
find significant differences in the color distribution as a function of the 
large-scale environment. 

These results, coupled with the observed differential galaxy clustering reported in 
\citet{Abbas_Sheth_06}, provide strong support to the hypothesis that an important 
fraction of the observed environmental effects on galaxy properties, like broadband 
optical color, or star formation history, are just the reflection of the correlation 
between galaxy stellar mass and the galaxy hosting dark matter halo mass, which in 
turn correlates with the surrounding large scale environment. Further discussion on 
the correlation between star formation activity and the large-scale environment where 
galaxies are located will be presented in a forthcoming paper (Vergani et al., 
in preparation).

\begin{acknowledgements}
MS would like to thank Frank van den Bosch for a very useful discussion on
environmental effects, that provided the motivation for this work.\\ 
This research has been developed within the framework of the VVDS consortium.
This work has been partially supported by the
CNRS-INSU and its Programme National de Cosmologie (France),
and by Italian Ministry (MIUR) grants COFIN2000 (MM02037133) and 
COFIN2003 (num.2003020150) and by INAF grants (PRIN-INAF 2005).
DV acknowledges the support through a Marie Curie ERG, funded by the European 
Commission under contract No. MERG-CT-2005-021704.\\
The VLT-VIMOS observations have been carried out on guaranteed
time (GTO) allocated by the European Southern Observatory (ESO)
to the VIRMOS consortium, under a contractual agreement between the
Centre National de la Recherche Scientifique of France, heading
a consortium of French and Italian institutes, and ESO,
to design, manufacture and test the VIMOS instrument.\\
Based on observations obtained with MegaPrime/MegaCam, a joint
project of CFHT and CEA/DAPNIA, at the Canada-France-Hawaii Telescope
(CFHT) which is operated by the National Research Council (NRC) of
Canada, the Institut National des Science de l'Univers of the Centre
National de la Recherche Scientifique (CNRS) of France, and the
University of Hawaii. This work is based in part on data products
produced at TERAPIX and the Canadian Astronomy Data Centre as part of
the Canada-France-Hawaii Telescope Legacy Survey, a collaborative
project of NRC and CNRS.

\end{acknowledgements}

\bibliographystyle{aa}
\bibliography{marcos.bib}

\end{document}